\documentstyle[aps,preprint,tighten,floats,epsf,rotate]{revtex}

\begin{document}
\draft

\preprint{\vbox{\it 
                        \null\hfill\rm    IP-BBSR/97-26, June'97}\\\\}
%
\title{Thermal Fluctuations of Disoriented Chiral Condensate Domains}
\author{Sanatan Digal \footnote{e-mail: digal@iop.ren.nic.in} and 
Ajit M. Srivastava \footnote{e-mail: ajit@iop.ren.nic.in}}
%
\address{Institute of Physics, Bhubaneswar 751005, India}
%
%
\maketitle
\widetext
\parshape=1 0.75in 5.5in
\begin{abstract}
 We argue that disoriented chiral condensate (DCC) domains are
not well defined for temperatures above the {\it Ginzburg}
temperature $T_G$ ($\simeq 0.7 T_c$). Above $T_G$, the dynamics of 
DCC domains is dominated by thermal fluctuations leading to fluctuating 
orientation of the chiral field in a given domain. It implies that DCC 
domains may form even in relatively lower energy collisions where the 
temperature only reaches $T_G$, and never rises to $T_c$. It also means 
that detection of DCC {\it can not} be taken as a signal for an 
intermediate chirally symmetric phase of matter. Using these 
considerations, we estimate the probability distribution for DCC 
domains as a function of the chiral angle.
\end{abstract}
\vskip 0.125 in
\parshape=1 -.75in 5.5in
\narrowtext

  Formation and detection of disoriented chiral condensates (DCC) in 
laboratory experiments has been a subject of intensive investigations
recently. By DCC one essentially means the formation of a chiral
condensate in an extended domain, such that the direction of the
condensate is misaligned from the true vacuum direction. It has been
suggested that regions of DCC may form in high multiplicity hadronic 
collisions or in relativistic heavy ion collisions 
\cite{anslm,bkt,blz,rw}. It has been argued that, as the chiral 
field relaxes to the true vacuum in such a domain, it will lead to 
coherent emission of pions which can be detected as anomalous 
fluctuations in neutral to charged pion ratio \cite{bkt}.

 The most important ingredient in most of the models, which have been 
proposed for DCC formation, is the assumption of an intermediate stage 
where chiral symmetry is restored (apart from the effects due to pion mass). 
Such a stage is naturally expected in the context of relativistic heavy 
ion collisions. In fact, DCC is thought to provide a relatively clean 
signal for an intermediate chirally symmetric phase of matter. Even 
in high multiplicity hadronic collisions such an intermediate stage may 
arise, though the thermalized region may not be large there. 

 In this paper, we consider thermal fluctuations of
correlation regions and its effects on the formation of DCC domains.
We argue that for temperatures below $T_c$ but above certain value
$T_G$ (the Ginzburg temperature), regions of correlation length size 
($\simeq m_\pi^{-1}$, say) can easily fluctuate to chirally symmetric
state, making  DCC domains ill defined.  A well defined DCC domain, inside 
which the field can evolve towards the true vacuum via field equations, 
can not exist until temperature has dropped below $T_G$ ($\simeq 0.7 T_c$
for $T_c$ in chiral limit, or about $0.8 T_c$ for $T_c$ with non-zero 
pion mass).  One implication of this result is that DCC domains can form 
even in relatively lower energy heavy-ion collisions where the temperature 
of the system never rises to $T_c$, but only rises upto $T_G$. 
Unfortunately, this also implies that detection of DCC, 
though interesting by itself, will not imply that quark-gluon plasma 
phase (or, more precisely, chirally symmetric 
phase of matter) has been detected. A hot hadronic gas, in the spontaneously 
broken phase of chiral symmetry, can still lead to formation of DCC as long 
as its temperature reaches a value near $T_G$.
 
 We first recall the basic picture of DCC formation in the framework of 
linear sigma model. We follow the notations used in ref. \onlinecite{kpst1}. 
The Lagrangian is expressed in terms of a scalar field 
$\sigma$ and the pion field \mbox{\boldmath $\pi$}.

\begin{equation}
\label{lagr}
{\cal L} \,=\, \frac{1}{2}\left(\partial_{\mu}\sigma\right)^2
+ \frac{1}{2}\left(\partial_{\mu} \mbox{\boldmath $\pi$}\right)^2
- \frac{\lambda}{4} \left(\sigma^2 + \mbox{\boldmath $\pi$}^2 -
c^2/\lambda\right)^2 + \epsilon \sigma \, .
\end{equation}

 Chiral symmetry is explicitly broken due to the last term $\epsilon \sigma$. 
The three parameters $\lambda$, $c$, and $\epsilon$, are restricted 
so as to give the proper pion mass, pion decay constant, a reasonable 
value for the $\sigma$ mass, PCAC (partial conservation of axial vector 
current), and the condition that the ground state of the theory occur 
at $\sigma = f_{\pi}$ and $\mbox{\boldmath $\pi$} = {\bf 0}$. We have,
 
\begin{eqnarray}
m_{\pi}^2 &=& \lambda f_{\pi}^2 - c^2 \, ,\\
m_{\sigma}^2 &=& 2 \lambda f_{\pi}^2 + m_{\pi}^2 \, ,\\
f_{\pi} m_{\pi}^2 &=& \epsilon \, .
\end{eqnarray}

 The values of physical parameters we take are, $m_{\pi}$ = 140 MeV, 
$f_{\pi}$ = 94.5 MeV, and $m_{\sigma}$ = 1 GeV. To obtain the effective 
potential, we expand the fields about an arbitrary point as

\begin{eqnarray}
\sigma(x) &=& v\,\cos\theta + \sigma'(x) \, ,\\
\mbox{\boldmath $\pi$}(x) &=& {\bf v}\,\sin\theta +
\mbox{\boldmath $\pi$}'(x) \, ,
\end{eqnarray}

where $v = |{\bf v}|$.  The primes denote fluctuations about the 
given point.  At one loop order, and in the high temperature 
approximation, the effective potential is

\begin{equation}
V(v, \theta; T) \,=\, \frac{\lambda}{4}v^4 - \frac{1}{2}
\left( c^2 - \frac{\lambda T^2}{2}
\right) v^2 - \epsilon v \cos\theta \, .
\end{equation}

Here $T$ is the temperature. In the chiral limit one finds, as is well
known, a second-order phase transition at the critical temperature
$T_c = \sqrt{2c^2/{\lambda}} = \sqrt{2} f_{\pi}$ ($\simeq 134$ MeV for 
our choice of parameters). When the non-zero quark masses are taken into 
account, chiral symmetry is explicitly broken, and is only approximately 
restored at a temperature $\simeq$ 115 MeV (in the sense that the saddle 
point at $\theta = \pi$ disappears above this temperature).  In the 
chiral limit, spontaneous breaking of chiral symmetry for temperatures
below $T_c$ implies that one particular point on the vacuum manifold 
$S^3$ (characterized by $\mbox{\boldmath $\pi$}^2 +  \sigma^2 = f_{\pi}^2$) 
will be chosen as the vacuum state in a given region of space, with all 
points on $S^3$ being equally likely. One 
may expect this to essentially hold true even in the presence of
(small) pion mass. This will lead to a sort of domain structure in 
the physical space where each domain will have the chiral field 
aligned in a given direction, but the directions in different domains 
vary randomly. 

 This essentially summarizes the physics of the formation of DCC domains. 
Formation of this type of domain structure in a phase transition has been 
extensively discussed in the context of topological defects in condensed 
matter physics and also in particle physics models of the early Universe 
\cite{top}. For an equilibrium, second order phase transition one expects 
\cite{rw} maximum size of domains to be of the order of $m_{\pi}^{-1}$.
For such small domains, dramatic signals like fluctuation in neutral to 
charged pion ratio will not be observable. [For small DCC domains, some 
other signals may be promising, such as, antibaryon enhancement \cite{kpst2}, 
enhancement in isospin violation \cite{iso}, etc.]. It was proposed
by Rajagopal and Wilczek \cite{rw} that large DCC domains may arise if 
the transition happens out of equilibrium. The idea is that the system 
starts in thermal equilibrium at high temperature where the vacuum 
expectation value of the field is near zero. After the quench, the field 
will find itself sitting at the top of the central bump of a low (say, zero)
temperature effective potential.  The field will then roll down with its 
dynamics governed by zero temperature field equations. It was shown in 
ref. \onlinecite{rw} that long wavelength modes can grow significantly during 
such a roll down.  However, it has been argued in ref. \onlinecite{gavin1} that 
due to strong coupling, the roll down of the field is very rapid, leading 
to small domains, perhaps 1-1.5 fm size.

 Large DCC domains may arise in the {\it annealing} scenario proposed
by  Gavin and M\"uller \cite{gavin2}. Here one uses the fact that it takes 
some time for the effective potential to change its shape, depending on 
the cooling rate of the plasma. Initially, at temperatures close to $T_c$,
the effective potential is nearly flat for $\sigma \simeq  \mbox{\boldmath 
$\pi$} \simeq 0$. Therefore, the roll down time scale can be very large at 
first. Only as the potential approaches its zero temperature shape does the 
roll down become rapid.  It was argued in ref.\onlinecite{gavin2} that 
domains as large as 5 fm may form in this scenario.

  We now turn to the consideration of thermal fluctuations of correlated 
domains. It is well known from the studies of phase transitions in 
condensed matter systems that, for second order transitions, thermal 
fluctuations of the order parameter become completely dominant for 
temperatures too close to the critical temperature \cite{tg1,tg2,tg3}. This 
is the reason that in discussions of the ordered phases in condensed matter 
systems, as well as in particle theory models of the early Universe (in 
the context of topological defect formation), the picture 
of correlation regions with well defined value of the order parameter
(i.e. the {\it domains}) makes sense only for temperatures below, what is 
known as, the Ginzburg temperature $T_G$, see ref. \onlinecite{top}. This 
temperature is defined as the one above which thermal fluctuations in the
value of the order parameter, about its mean value, become much larger
than the mean value itself \cite{tg1}. Equivalently, one can think of 
$T_G$ as the temperature above which a domain of correlation length size can 
easily fluctuate to the symmetric phase by thermal effects \cite{tg2,top}. 
Only when such fluctuations become sufficiently suppressed, can one talk 
about a well defined domain with the order parameter field taking well 
defined orientation inside the domain (with subsequent evolution via
field equations, etc.). This is why, one usually talks about the formation
of topological defects via domain structure only for temperatures 
below $T_G$, see ref. \onlinecite{top}. [Though, there are subtle points
here as far as large scale structure of defects is concerned, see
ref. \onlinecite{zurk}.]

   An intuitive way of estimating the value of $T_G$ is as follows. $T_G$
can be obtained by equating the temperature to the energy $E_d$ needed to
fluctuate a correlation domain to the chirally symmetric phase. $E_d$ is 
equal to the energy density difference $\bigtriangleup V(T)$ between the 
bottom of the effective potential and the top of the central bump, times 
the volume of a region of size correlation length. Here, $m_\sigma^{-1}$ 
should determine the appropriate correlation length \cite{top,tg1,tg2,tg3}. 
Though, for a conservative estimate of $T_G$, we will take $m_\pi^{-1}$ 
(as this is the largest length scale). Thus we get,

\begin{equation}
E_d \equiv m_\pi^{-3}(T) \bigtriangleup V(T) \simeq T ~ .
\end{equation}

 Here $m_\pi(T)$ is the temperature dependent pion mass \cite{rw}, 
and $\bigtriangleup V(T)$ is calculated from Eq.(7). We mention here 
that we are neglecting any gradient energy terms in this estimate 
of $E_d$, as we are only interested in rough estimate 
of $T_G$. Also, if we take the correlation length to be given by 
$m_\sigma^{-1}$ (which is more appropriate) in the above equation, then $T_G$ 
comes out to be extremely small $\simeq 2 - 4$ MeV. 
Clearly such a small value of $T_G$ will have drastic implications. 
We should point out that this $T_G$ will be larger (though still much 
less than the value obtained from Eq.(8)) if gradient terms are also 
incorporated in Eq.(8). [In fact, the gradient term can make the 
orientational fluctuations of the chiral field in a small domain 
to be energetically unfavorable compared to the magnitude fluctuations. 
Though such estimates of orientational fluctuations may not be entirely 
justifiable, pion being the (approximate) Goldstone boson \cite{tg3}.] At 
this preliminary stage, we will not worry about these complications as they 
do not seem to have too significant effect on the estimates of $T_G$ obtained 
from Eq.(8). The main aim of this paper is to explore the physical 
consequences of the considerations of thermal fluctuations with the 
estimate of $T_G$ which is significantly lower than $T_c$. We hope to 
present a more detailed calculation of $T_G$ in a future work.

 An interesting aspect for the case of chiral transition is that the bottom 
of the potential is tilted. Thus the value of $T_G$, determined by solving 
Eq.(8) for $T$, will depend on the chiral angle $\theta$. Figure 1 shows 
the plot of $T_G$ as a function of the chiral angle $\theta$.  [Though, we
mention that the true value of $T_G$, in the sense of defining the regime
where fluctuations are dominant, corresponds to the value of $T_G$ at
$\theta = \pi$ in Fig.1. The plot of $T_G$ vs. $\theta$ is in the context 
of DCC formation, and shows the values of temperature at which domains
with various values of $\theta$ become unstable due to thermal 
fluctuations.]  

   There are two important points to be noted here. First, note that 
these values of $T_G$ are significantly lower than the value of $T_c$,
especially for large $\theta$. For $\theta = \pi$, $T_G \simeq 0.6 T_c$.
For heavy-ion collisions, in the longitudinal expansion 
model of Bjorken \cite{bj}, the temperature 
of the expanding fluid is proportional to $\tau^{-1/3}$, $\tau$ being the
proper time. If we take that $T = T_c$ is achieved at $\tau \simeq 3 - 10$
fm, then the temperature will drop to $0.6 T_c$  at $\tau \simeq 14 - 46$ fm. 
These values of $\tau$ are large from the point of view of the expected life 
time of the hadronic gas before freeze out. Also, note that in the context 
of the annealing scenario of DCC formation \cite{gavin2}, for temperatures 
as low as $0.6 T_c$, the bottom of the effective potential is not that flat 
any more (compared to the situation close to $T_c$), and the roll down of
the chiral field will be much faster. 
The net result will be that the resulting DCC domain will not be as large.  
For, example, $m_\sigma$ may grow \cite{rw} by about a factor four as 
temperature drops from $T_c$ to 0.6$T_c$, leading to DCC domains  which
are smaller by a factor of two or so, say of size 2 - 2.5 fm.

 Second point is that, due to the dependence of $T_G$ on $\theta$, 
domains of smaller $\theta$ will become stable first (against thermal
fluctuations) and domains with larger values of $\theta$ will stabilize 
later. [For $\theta$ between 0 and $\pi$. Things being symmetric about 
$\theta = \pi$.]
During this time interval, a region with large value of $\theta$ 
will still keep fluctuating to chirally symmetric phase (as the temperature 
is still above $T_G$ relevant for that value of $\theta$).  During these 
fluctuations, the magnitude of the chiral field is fluctuating between zero 
and the value corresponding to the broken phase.  This then leads to 
fluctuating orientation of the chiral field in that region, with decreasing 
probability for the region to retain its large value of $\theta$, as 
$\theta$ in that region can take any value after each fluctuation. 
[For large $\theta$ one may have to worry about the gradient energy terms. 
However, as for Eq.(8), we will not worry about it for the present rough
estimates.] This leads 
to a probability distribution for different $\theta$ values, with smaller 
values of $\theta$ being most probable and large values of $\theta$
being suppressed. We now calculate the time evolution of this probability
distribution. For simplicity we discretize $\theta$ between 0 to $\pi$
in $m$ segments (we will use $m = 6$). Rate of change of the number 
of DCC domains $N_i(\tau)$ with value of the chiral angle in the ith 
segment, at proper time $\tau$, can be written as,

$${dN_i(\tau) \over d\tau} = - A~exp[-E_d(i,T)/T]~ ({m - 1 \over m})~ N_i $$  
\begin{equation}
 + ~ \sum_{j, j \ne i} A~exp[-E_d(j,T)/T]~ {1 \over m} ~ N_j ~ .
\end{equation} 

 Here, the factors $(m-1)/m$, and  $1/m$, give probabilities for fluctuation 
of a domain out of, and into, the ith segment respectively. We will only be 
interested in rough estimates of the relative values of 
$N_i(\tau)$, so we drop the pre-exponential factor $A$ for the thermal 
fluctuation rate in the above equation (assuming that $A$ does not depend 
on $\theta$). $E_d(i,T)$ denotes the value of $E_d$ (Eq.(8)) for the
domain with $\theta$ in the ith segment. For the time evolution of
the temperature we assume longitudinal expansion model of Bjorken
\cite{bj} for the plasma,

\begin{equation}
T(\tau) = T(\tau_0) ({\tau_0 \over \tau})^{1/3} ~ .
\end{equation}

 We take $\tau_0 =  3$ fm as an example, corresponding to the chiral
symmetry breaking transition, with $T_0 = T_c \simeq 134$ MeV.
Initial values of $N_i$ are prescribed at a temperature $T \simeq 110$ 
MeV (at $\tau \simeq 5$ fm), when the central bump has already appeared 
in the effective potential in Eq.(7) (see, ref. \onlinecite{kpst1}). Figure 2 
shows the plots of $N(\tau)$. [We mention again that the plots in Fig.2 
are only supposed to represent relative values of $N_i$s, as we 
dropped the pre-exponential factor $A$ in Eq.(9).] 
 
 We see from Fig.2 that the numbers of DCC domains with different values of 
$\theta$ rapidly diverge from each other, approaching somewhat constant 
values asymptotically. In fact, the memory of initial condition seems to be 
lost rapidly and curves of a given $\theta$ approach approximately same 
asymptotic value for both of the sets corresponding to two different initial
conditions. This shows the robustness of the distribution of DCC domains with 
different $\theta$ values. Of course, at later times, when thermal 
fluctuations are suppressed, classical evolution of the field towards the 
true vacuum will take over, enhancing the number of domains with $\theta = 0$ 
and depleting others. We neglect any such evolution as our aim is only to 
provide a reasonable estimate of relative populations of DCC domains with
different $\theta$, which can then be taken as the initial condition
to study the roll down of the field and domain growth etc. Around
at $\tau \simeq 16$ fm (which corresponds to $T \simeq 0.6 T_c$)
we find that ratio between number of DCC domains with $\theta \simeq
\pi$ and $\theta \simeq 0$ is about 0.3. A nontrivial feature of
these plots is that for all values of $\theta$, the asymptotic value
of $N(\tau)$ remains non-zero, with $N(\tau)$ for $\theta \simeq \pi/2$
approaching a value appropriate for a uniform $\theta$ distribution.
 
  This brings us to another very important aspect of these considerations
of thermal fluctuations of DCC domains. As domains keep fluctuating, and 
keep re-distributing $\theta$, for any temperature not too low compared 
to $T_G$, it gives a way to {\it disoreint} the chiral vacuum without ever 
achieving an intermediate stage where the chiral symmetry is restored. 
Conventionally, in most approaches, such a stage is believed to be essential 
for disorienting the chiral vacuum. However, our considerations show that
DCC domains can form even if the temperature of the hadronic matter, 
in heavy ion collisions or in hadronic collisions, only reaches a value 
near $T_G$ ($\simeq 0.7 T_c$ for our parameters) and never rises above 
$T_c$. This means that DCC formation may be accessible even in experimental 
situations with relatively lower center of mass energy.

  However, at the same time, one is forced to conclude that detection of
DCC (via any of the signals which have been proposed in the literature)
{\it will not} imply that a new phase of the hadronic matter has been
detected in which chiral symmetry is restored. DCC domains can form
in the hadronic matter in the chiral symmetry broken phase itself, as
long as its temperature reaches a value near $0.7 T_c$ (corresponding 
to $T_G$ for $\theta = 0$). In some 
sense the chirally symmetric state would have been {\it probed} by the 
thermal fluctuations leading to  DCC formation at such low temperatures. 
Still, reaching a different phase of the matter in the thermodynamic 
sense is not required. Certainly, detection of DCC will be interesting 
in itself as it probes the non-perturbative aspects of the chiral model,
and may provide explanation for the Centauro events \cite{bkt}.

  We summarize by emphasizing the main points of our results. We have
argued that thermal fluctuations of correlation regions make DCC
domains ill defined until the temperature of the hadronic system
drops to the Ginzburg temperature $T_G$ ($\simeq 0.7 T_c$ for our 
choice of parameters). This means that studies of domain growth 
and the roll down of the chiral field towards the true vacuum etc.
via field equations, can be performed only at temperatures lower than 
$T_G$. By considering thermal fluctuations and resulting re-distribution
of $\theta$ in different DCC domains, we estimate how relative proportions
of DCC domains with different values of $\theta$ change in time.
Our results imply that DCC formation may be accessible even in relatively
lower energy collisions. At the same time it implies that DCC can not
be thought of as a signal for detecting an intermediate, chirally
symmetric, phase of matter.


 We are very thankful to Pankaj Agrawal, Shashi Phatak, and Supratim 
Sengupta  for many useful discussions and comments.

\vspace*{8mm}
\centerline{\bf FIGURE CAPTIONS}
\vspace*{4mm}

 Fig.1: Ginzburg temperature $T_G$ (signifying the onset of 
fluctuations for a domain with given value of $\theta$) as a function of 
the chiral angle $\theta$ (in degrees) for $T_c$ = 134 MeV. 

 Fig.2: Evolution of the relative proportion of DCC domains with
different values of the chiral angle $\theta$. The set of curves 
with solid lines denotes a uniform initial distribution of $\theta$ 
with $N_i = 10, i = 1,..6$, while the set of dashed curves corresponds 
to the initial condition with $N_1$ (corresponding to $\theta = 
0^0 - 30^0$ segment) equal to 60 while all 
other $N_i$s are zero initially. For both the 
sets, curves from the top to the bottom correspond to values of 
$\theta = 15^0, 45^0, 75^0, 105^0, 135^0,$ and $165^0$.

\newpage

\begin{figure}[h]
\begin{center}
\leavevmode
\epsfysize=20truecm \vbox{\epsfbox{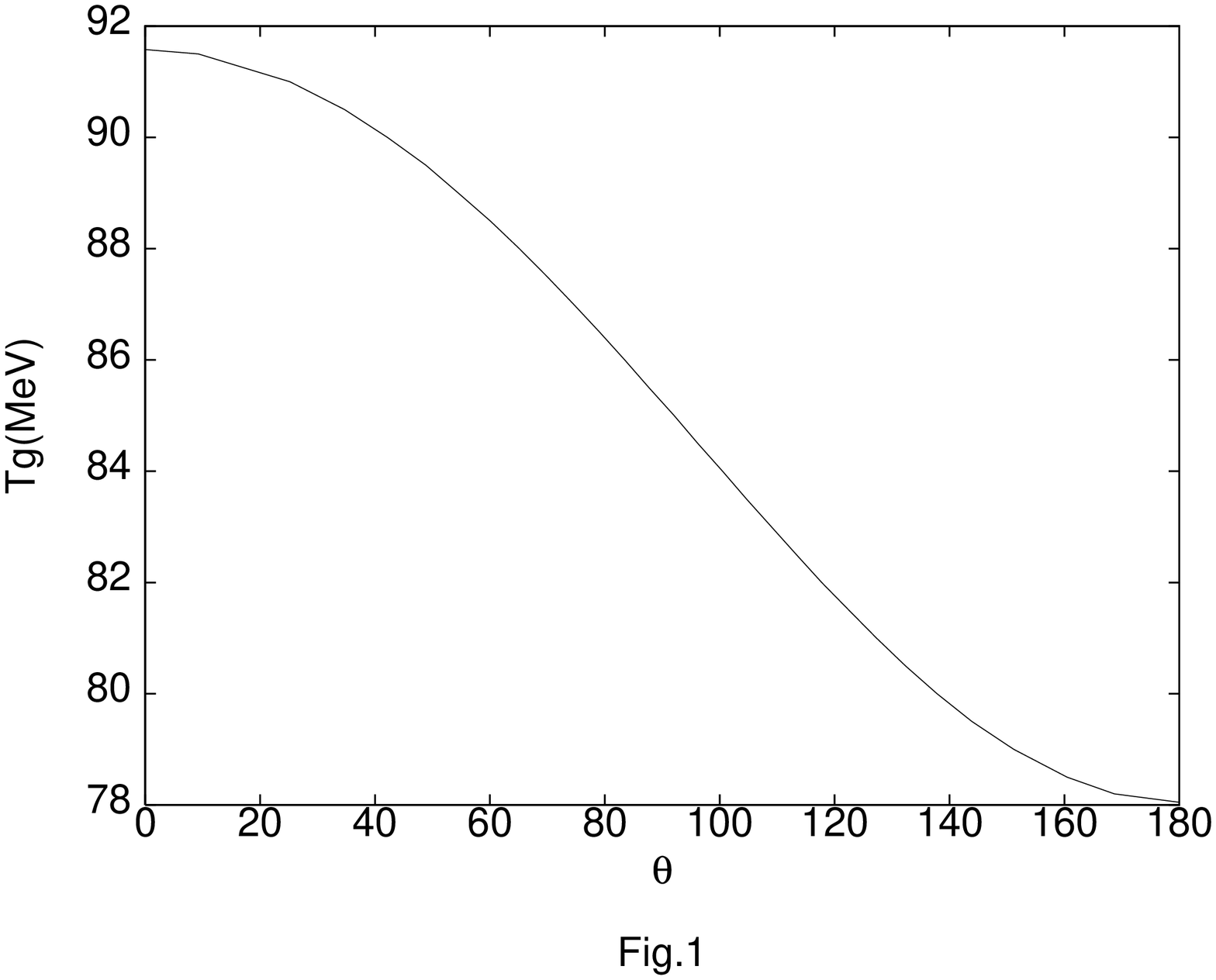}}
\end{center}
\end{figure}

\newpage

\begin{figure}[h]
\begin{center}
\leavevmode
\epsfysize=20truecm \vbox{\epsfbox{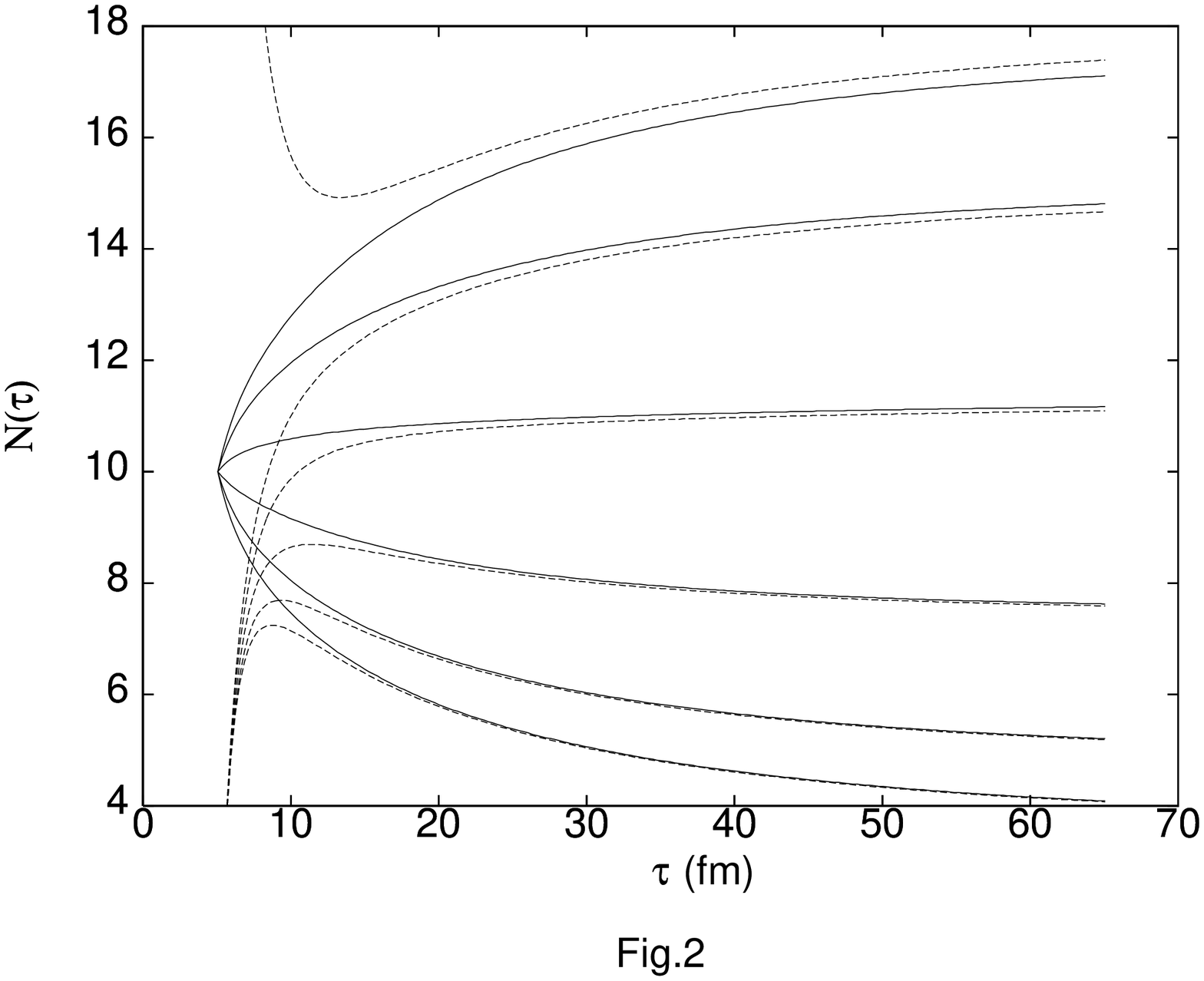}}
\end{center}
\end{figure}

\end{document}